\documentclass[a4paper,11pt]{article}	
\pdfoutput=1 
\usepackage{jheppub} 

\usepackage[table]{xcolor}                  
\usepackage{colortbl}
\usepackage{booktabs}                       
\usepackage{multirow,bigdelim}              
\usepackage{longtable}                      
\usepackage{arydshln}                       

\usepackage{units}                          
\usepackage{amsmath,amssymb,mathtools}      
\usepackage{slashed}                        
\usepackage{amsthm}
\usepackage{bbm,dsfont}                     

\usepackage{bm}                             
\usepackage{upgreek}                        
\usepackage[vcentermath]{youngtab}          
\usepackage{ytableau}                       
\usepackage{empheq}                         
\usepackage{suffix}	                        
\usepackage{lscape}	                        

\usepackage{graphicx}                       
\usepackage{subfig}	                        
\usepackage{tikz}                           
\usepackage{float}
\usepackage[export]{adjustbox}              

\usepackage{url}                            
\usepackage{makeidx}                       
\usepackage[numbers,sort&compress]{natbib}  
\usepackage[colorlinks=true,urlcolor=blue,anchorcolor=blue,citecolor=blue,filecolor=blue,linkcolor=blue,menucolor=blue,pagecolor=blue,linktocpage=true,pdfproducer=medialab,pdfa=true]{hyperref}	                    

\numberwithin{equation}{section}        
\numberwithin{table}{section}
\numberwithin{figure}{section}

\graphicspath{{figs/}}

\usepackage{makeidx}

\makeindex

\newcommand{\und}[1]{\underline{#1}}

\newcommand{\ep}{\epsilon}

\newcommand{\dDe}[1]{\hspace{-0.5ex}
  \raisebox{.23ex}{
  {$\stackrel{\raisebox{-.23ex}{$\scriptscriptstyle\bullet$}}\Delta_{
  \raisebox{-.23ex}[1ex][0ex]{$\scriptstyle#1$}}$}}
}

\def\mc{\mathcal}

\def\mf{\mathfrak}
\def\ms{\mathsf}

\newcommand{\wh}[1]{\widehat{#1}}
\newcommand{\pd}{\partial}

\newcommand{\re}{{\rm e}}
\newcommand{\ri}{{\mathsf{i}}}
\newcommand{\rd}{{\rm d}}

\newcommand{\nn}{\nonumber \\}

\DeclareMathOperator{\imag}{Im}
\def\vev#1{\left\langle #1 \right\rangle}

\def\IC{\mathbb{C}}

\def\IF{\mathbb{F}}

\def\IP{\mathbb{P}}

\def\IR{\mathbb{R}}
\def\IZ{\mathbb{Z}}

\newcommand{\CA}{\mc{A}}
\newcommand{\CC}{\mc{C}}
\newcommand{\CF}{\mc{F}}
\newcommand{\CG}{\mc{G}}
\newcommand{\CO}{\mc{O}}
\newcommand{\CM}{\mc{M}}
\newcommand{\CV}{\mc{V}}
\newcommand{\CW}{\mc{W}}

\newcommand{\CAp}{\CA_{(\und{p},\und{q},\und{s})}}

\newcommand{\gt}{{t^\Gamma}}

\newcommand{\gCF}{{\CF^\Gamma}}
\newcommand{\gC}{{C^\Gamma}}
\newcommand{\gLambda}{{\Lambda^\Gamma}}
\newcommand{\gR}{{R^\Gamma}}

\newcommand{\gti}{t^{\Gamma,i}}
\newcommand{\gtDl}{t^\Gamma_{D,\ell}}
\newcommand{\gTheta}{{\Theta^\Gamma}}

\newcommand{\un}{\und{n}}
\newcommand{\um}{\und{m}}
\newcommand{\uq}{\und{q}}

\newcommand{\ur}{\und{r}}
\newcommand{\ut}{{\und{t}}}
\newcommand{\uC}{\und{C}}
\newcommand{\ugt}{{\und{t}^\Gamma}}
\newcommand{\ugR}{{\und{R}^\Gamma}}
\newcommand{\ugC}{{\und{C}^\Gamma}}
\newcommand{\uCC}{{\und{\mc{C}}}}
\newcommand{\ugCC}{{\und{\mc{C}}}^\Gamma}
\newcommand{\ugr}{{\und{r}^\Gamma}}
\newcommand{\uz}{{\und{z}}}
\newcommand{\uR}{{\und{R}}}
\newcommand{\uB}{{\und{B}}}

\newcommand{\SSATop}{\ms{S}_{\CA}^{\text{top}}}
\newcommand{\SSkATop}{\ms{S}_{k\CA}^{\text{top}}}
\newcommand{\SSANS}{\ms{S}_{\CA}^{\text{NS}}}

\title{\boldmath Relations between Stokes constants of unrefined and
  Nekrasov--Shatashvili topological strings}%
\author[a]{Jie Gu}
\affiliation[a]{School of Physics and Shing-Tung Yau Center\\
  Southeast University, Nanjing 210096, China}%

\emailAdd{eij.ug.phys@gmail.com}

\abstract{In this paper we demonstrate that the Stokes constants of
  unrefined free energies and the Stokes constants of
  Nekrasov--Shatashvili free energies of topological string on a
  non-compact Calabi--Yau threefold are identical, possibly up to a
  sign, for any Borel singularity which is not associated to a compact
  two-cycle that intersects only with non-compact four-cycles.  Since
  the Stokes constants of Nekrasov--Shatashvili free energies are
  conjectured to coincide with those of quantum periods and therefore
  have the interpretation of BPS invariants, our results give strong
  support that the Stokes constants of unrefined free energies may
  also be identified with BPS invariants.}

\keywords{Topological string, resurgence, Borel--Laplace resummation,
  Stokes automorphisms, Stokes constants, unrefined free energies,
  Nekrasov--Shatashvili free energies, BPS invariants, blowup
  equations, quantum periods, Wilson loops}

\begin{document}
\maketitle
\flushbottom

\section{Introduction}

String theory has rich non-perturbative structures. For instance, it
was already pointed out in \cite{Gross:1988ib} for bosonic string and
later in \cite{Shenker:1990uf} for fermionic string that the free
energy as a perturbative series in string coupling $g_s$ is divergent
as the coefficients have factorial growth, i.e.
\begin{equation}
  \label{eq:Fgfac}
  \CF(g_s) = \sum_{g=0}^\infty \CF_g g_s^{2g-2},\quad \CF_g\sim (2g)!,
\end{equation}
which implies the power series has zero radius of convergence.  Such a
factorial growth of coefficients also signals that the full and exact
free energy would require exponentially small non-perturbative
corrections.  These non-perturbative corrections were interpreted as
effects from D-branes \cite{Polchinski:1994fq}, although detailed
calculations of D-branes were only performed recently
\cite{Alexandrov:2003nn,Eniceicu:2022dru} to reproduce the
exponentially small non-perturbative corrections in the cases of
minimal string theory.  In general the non-perturbative corrections to
string free energies are difficult to study.

One division of string theory where one might have a better chance of
understanding the non-perturbative corrections is topological string
theory.  Topological string is constructed by topological twists on
the worldsheet of perturbative string, and it captures the important
BPS sectors of type II superstring compactified on a Calabi--Yau
threefold.  On the other hand, topological string is relatively
simple. It permits a rather rigorous mathematical definition through
the Gromov-Witten theory, and the perturbative free energies in
topological sting can be computed via various methods, including
holomorphic anomaly equations
\cite{Bershadsky:1993cx,Bershadsky:1993ta,Huang:2011qx}, topological
vertex \cite{Aganagic:2003db,Iqbal:2007ii}, topological recursion
\cite{Bouchard:2007ys}, blowup equations \cite{Huang:2017mis}, and
etc.
Making use of these methods, hundred of terms of perturbative free
energies can be computed in the case of non-compact Calabi--Yau
threefolds, and more than sixty terms have recently been computed for
certain family of compact Calabi--Yau threefolds in
\cite{Alexandrov:2023zjb}, building on \cite{Huang:2006hq}, including
the famous quintic model, which provides a treasure of perturbative
data unimaginable in critical string theory, making tests of various
proposals of non-perturbative corrections possible.  It is due to these
facts that it was mused in ``A panorama of physical mathematics
c.~2022'' \cite{Bah:2022wot} that if we wish to make progress on
understanding what is string theory including all of its
non-perturbative aspects, it is more tractable and more realistic to
try to answer this question first in topological string.

Among different proposals to study non-perturbative corrections to
topological string free energies, one of the most promising is to use
the resurgence theory \cite{Ecalle}\footnote{See lectures by both
  mathematicians and physicists
  \cite{Marino:2012zq,Mitschi:2016fxp,Aniceto:2018bis}}.  The key idea
is that non-perturbative corrections to a divergent perturbative
series appear in the form of trans-series, and the full and exact
solution can be written as the Borel--Laplace resummation, a
distingshed method to resumm a divergent power series, of certain
linear combination of these trans-series in addition to the
perturbative series.  Furthermore, in order for the resummed linear
combination to be a well-defined function, the trans-series
representing non-perturbative corrections must be encoded in and
therefore can be extracted from the perturbative series, through
various Stokes automorphisms characterised by a collection of numbers
called Stokes constants.  The resurgence theory therefore offers a
roadmap to systematically study non-perturbative contributions to
topological string free energy, making use of the already available
rich data of perturbative expansion.

Resurgence techniques were first applied to study topological string
in \cite{Marino:2006hs,Marino:2007te,Marino:2008ya}, where for
instance it was checked numerically that indeed topological string
free energies grow like \eqref{eq:Fgfac}, and the non-perturbative
effects that control such a growth behavior were analysed in simple
models.  Later studies focused on the simplest model, the resolved
conifold, where both the trans-series and the Stokes constants could
be computed \cite{Pasquetti2010} (see also
\cite{Alim:2021mhp,Alim:2022oll,Grassi:2022zuk} building on
\cite{Pasquetti2010}).  For more complicated geometries, these
computations are much more difficult.  Nevertheless, based on a novel
method proposed in \cite{Couso-Santamaria2016,Couso-Santamaria2015} to
systematically calculate the trans-series using holomorphic anomaly
equations \cite{Bershadsky:1993cx,Bershadsky:1993ta}, the trans-series
in any non-perturbative sectors were calculated in closed form
\cite{Gu:2022sqc,Gu:2023mgf} for both compact and non-compact
Calabi--Yau threefolds.  In a different line of researches, Tom
Bridgeland proposed that the BPS invariants or generalised DT
invariants of a non-compact Calabi--Yau threefold can be used to
construct a Riemann-Hilbert problem whose solution of $\tau$ function
expands to the topological string free energy
\cite{Bridgeland:2016nqw,Bridgeland:2017vbr}.
These BPS invariants are countings of stable D-brane bound states.
Various subsets of BPS invariants are already known to be related to
topological string.  For instance, the Gopakumar-Vafa formula
\cite{Gopakumar:1998ii,Gopakumar:1998jq} expresses the perturbative
free energy in terms of the D2-D0 BPS invariants, and the generating
function of D6-D2-D0 BPS invariants, also known as the PT invariants,
is also related to the topological string free energy
\cite{Maulik:2003rzb}.  Bridgeland's proposal provides another
interesting link between basically Stokes constants of topological
string free energies and BPS invariants, but concrete construction
could only be made for the resolved conifold (see also
\cite{Alim:2021mhp}), where the only non-trivial BPS invariants are
countings of D2-D0 bound states.  Various hints exist for more
complicated models, for instance checks of integrality of Stokes
constants in \cite{Gu:2021ize,Gu:2022sqc,Gu:2023mgf} in certain
models, but no conclusive statements can be made, and many questions
still remain open, including: if the relation between Stokes constants
and BPS invariants can be generalised, and if true, what are the
concrete formulas, and whether the full set or only a subset of BPS
invariants can be recovered from the Stokes constants.

\begin{figure}
  \centering
  \includegraphics[width=0.6\linewidth]{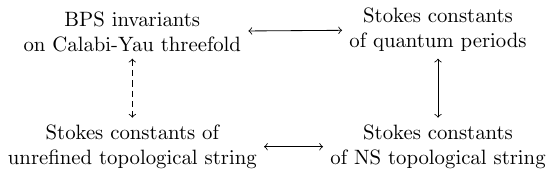}
  \caption{The relations between BPS invariants, Stokes constants of
    quantum periods, of NS topological string, and of unrefined
    topological string.}
  \label{fig:summary}
\end{figure}

In this short note, we will make an important step towards answering
these questions through a slightly different but related route,
summarised by Fig.~\ref{fig:summary}.  When the target space is a
suitable noncompact Calabi--Yau threefold $X$, topological string is
characterised by a Riemann surface called the mirror curve equipped
with a canonical 1-form associated to $X$ by mirror symmetry.  At the
same time, it is equivalent to a 5d $N=1$ gauge theory on
$S^1\times \IR^4$, which reduces to an $N=2$ gauge theory when $S^1$
shrinks to a point, corresponding to certain scaling limit of the
mirror curve \cite{Katz:1996fh}.  It was found in \cite{Klemm:1996bj}
that the canonical 1-form describes a metric on the mirror curve, and
BPS states of either the topological string or the supersymmetric
field theory can be described by geodesic 1-cycles on the mirror curve
with respect to this metric.  This observation was developed into a
full-blown method to calculate BPS invariants in
\cite{Gaiotto:2009hg} known either as the spectral network
(4d field theory) \cite{Gaiotto:2009hg,Gaiotto:2012rg} or
exponential network (5d field theory)
\cite{Banerjee:2018syt,Banerjee:2019apt,Banerjee:2020moh}.

On the other hand, it was pointed out that the mirror curve can be
promoted via quantisation to either a differential operator (4d field
theory) or a difference operator (5d field theory), called the quantum
mirror curve, and the periods of the mirror curves are promoted to
quantum periods which are divergent power series in $\hbar$ through
the exact WKB method.  These quantum periods have remarkable resurgent
properties.  When the quantum mirror curve is a second order
differential operator (4d rank one gauge theory), it was proved that
the trans-series that appear in Stokes automorphisms of quantum
periods are associated also with geodesic 1-cycles
\cite{Dillinger:1993} and they should be naturally mapped to the BPS
states.  Furthermore, the Stokes automorphism takes a form, known as
the Delabaere--Dillinger--Pham formula
\cite{Delabaere1999,Delabaere19971:exact,Dillinger:1993}, which
resembles the Kontsevich-Soibelman automorphism, a crucial ingredient
of the wall-crossing formula
\cite{Kontsevich:2008fj,Kontsevich:2010px}, also known as the spectrum
generator of BPS invariants \cite{Gaiotto2008cd}.  The Stokes
constants, which are coefficients of the Stokes automorphisms, should
then be identified with the BPS invariants, which are coefficients of
the Kontsevich-Soibelman automorphisms.  This identification was
checked in various 4d rank one gauge theories
\cite{Grassi:2019coc,Grassi:2021wpw} and is expected to hold in higher
rank theories.  This provides the upper horizontal arrows in
Fig.~\ref{fig:summary}.

Quantum periods can be studied through another approach different from
the exact WKB method.  Inspired by Nekrasov's partition function for
4d and 5d gauge theories on the Omega background
\cite{Nekrasov:2002qd}, the topological string free energy was refined
to depend on not a single perturbative expansion parameter $g_s$ but
two $\ep_1,\ep_2$, and it reduces to the unrefined case in the limit
$\ep_1+\ep_2 = 0$.  In another special limit $\ep_2\rightarrow 0$,
called the Nekrasov--Shatashvili limit \cite{Nekrasov:2009rc}, the
topological string free energy provides a set of relations between
quantum A- and B-periods called the quantum special geometry
\cite{Aganagic:2011mi}.  The Wilson loop amplitudes in various gauge
representations in the NS limit then provide another set of relations
between these quantum periods.  Together the NS free energies and the
NS Wilson loop amplitudes completely determine the quantum periods
\cite{Gu:2022fss}, and the Stokes automorphisms of quantum periods can
also be derived from those of NS free energies and Wilson loop amplitudes.  A
pattern of the latter was recognised, from which one concludes that
the Stokes automorphisms of quantum periods in 5d gauge theories
follow the same DDP formulas as 4d gauge theories \cite{Gu:2022fss}
(see also \cite{DelMonte:2022kxh}), further confirming the horizontal
arrows in the second arrow in Fig.~\ref{fig:summary}.  In addition,
Stokes constants of NS free energies are themselves identified with
Stokes constants of quantum periods, providing the vertical arrows on
the right hand side in Fig.~\ref{fig:summary}.

In this note, we will argue that the Stokes constants of unrefined
free energies and those of NS free energies of topological string can
also be identified, possibly up to a sign, given by the key formula
\eqref{eq:TopNSRel}, thus providing the lower horizontal arrows in
Fig.~\ref{fig:summary}.  The key idea is to use the blowup equations
for refined topological string free energy
\cite{Gu:2017ccq,Huang:2017mis}, which in a special limit provides the
sought for relationship between unrefined and NS free energies known
as the compatibility formula
\cite{Sun2016,Grassi:2016nnt,Huang:2017mis}.  Once the lower
horizontal arrows are established, one can make direct identification
betwene Stokes constants of unrefined topological string free energy
and BPS invariants of the Calabi--Yau threefold, given by
eq.~\eqref{eq:SSTopBPS}, represented by the dashed vertical arrows on
the left hand side in Fig.~\ref{fig:summary}.  The identified BPS
invariants include all the D4-D2-D0 stable bound states.

We emphysize that our argument only works on topological string on
non-compact Calabi--Yau threefolds where NS free energies and quantum
periods as intermediary steps can be defined.  A direct argument that
make the left vertical arrow that works also for compact Calabi--Yau
threefolds would be very desirable, possibly along the line of
\cite{Iwaki:2023rst}.

The remainder of the paper is structured as follows.  In section
\ref{sc:resurg-top}, we sketch the ingredients of resurgence theory
enough for understanding the derivation and the statements in this
paper.  We then summarise the known results on Stokes automorphisms
for both unrefined free energies \cite{Gu:2022sqc,Gu:2023mgf} and NS
free energies \cite{Gu:2022fss}.
In section \ref{sc:blowup}, we introduce and slightly extend the
blowup equations for refined topological string, which are important
for later sections.  We give the key argument in section
\ref{sc:TS-NS} that makes the connection between the Stokes constants
of unrefined and NS free energies.  Finally, we summarise and discuss
open problems in section \ref{sc:disc}.  In Appendix \ref{sc:qPerds},
we re-derive and generalise the relationship between Stokes constants
of NS free energies and those of quantum periods, which is crucial for
closing the circle in Fig.~\ref{fig:summary}.

\section{Resurgence and topological string}
\label{sc:resurg-top}

\subsection{Resurgence theory in a nutshell}

We give a lightning overview of the resurgence theory \cite{Ecalle}.
We refer to the lectures
\cite{Marino:2012zq,Mitschi:2016fxp,Aniceto:2018bis} for details.
Suppose we have a perturbative series $\varphi(z)$ of the Gevrey-1
type, i.e.
\begin{equation}
  \varphi(z) = \sum_{n\geq 0} a_n z^n,\quad a_n \sim n!,
\end{equation}
which is divergent with zero radius of convergent, the resurgence
theory tells us that if we wish to find the full and exact description
of the quantity in the form of a function $f(z)$ that admits
$\varphi(z)$ as an asymptotic expansion, we must include
non-perturbative corrections, which are actually encoded in and
therefore can be extracted from the perturbative series itself.

In order to uncover the hidden non-perturbative corrections, we
introduce the Borel transform,
\begin{equation}
  \wh{\varphi}(\zeta) = \sum_{n\geq 0}\frac{a_n}{n!}\zeta^n.
\end{equation}
This is a convergent series with a positive radius of convergence in
the complex $\zeta$-plane $\IC_\zeta$, also known as the Borel plane,
and it can be analytically continued to the entire complex plane with
possible singularities, known as the Borel singularities.  If the
singularities are discrete points and form a closed subset
$\Omega\subset \IC_\zeta$ and in addition $\wh{\varphi}(\zeta)$ allows
analytic continuation along any path in $\IC_\zeta\backslash \Omega$,
$\wh{\varphi}(\zeta)$ is called a resurgent function, and $\varphi(z)$
called a resurgent series.  In this case, we can define the Laplace
transform of $\wh{\varphi}(\zeta)$ along any direction which does not
pass through any Borel singularities\footnote{We also need to assume
  that $\wh{\varphi}(\zeta)$ has at most exponential growth
  $\sim \re^{|\zeta|/R}$ with $R>|z|$ when
  $\zeta \rightarrow \infty$.}
\begin{equation}
  \label{eq:BLsum}
  s_\theta(\varphi)(z) =
  \frac{1}{z}\int_0^{\re^{\ri\theta}\infty}\re^{-\zeta/z}\wh{\varphi}(\zeta)
  {\rd }\zeta.
\end{equation}
with $\theta = \arg z$.  This is known as the Borel--Laplace
resummation of $\varphi(z)$.

\begin{figure}
  \centering
  \includegraphics[height=5cm]{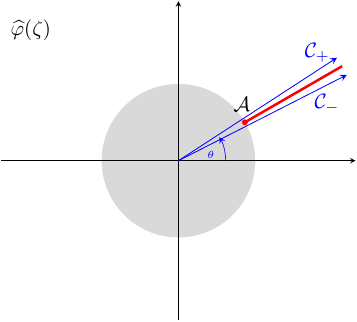}
  \caption{The lateral Borel resummations which sandwich a Stokes ray
    that pass through the singular point $\CA$.}
  \label{fig:latBorel}
\end{figure}

According to the resurgence theory, each of the discrete singular
points in the Borel plane in fact represents a non-perturbative saddle
point\footnote{They could also be renormalons which have no
  semi-classical saddle point interpretation.  But this distinction is
  irrelevant for our discussions.}  whose action is given by the
position $\CA$ of the singular point, and the perturbative series
$\varphi^{(\CA)}(z)$ in the non-perturbative sector can be uncovered
by the remarkable formula
\begin{equation}
  \label{eq:disc}
  s_{\theta_+}\varphi(z) - s_{\theta_-}\varphi(z) = \ms{s}_\CA
  \re^{-\CA/z} s_{\theta_-}\varphi^{(\CA)}(z).
\end{equation}
Here we change \eqref{eq:BLsum} slightly and define lateral
Borel--Laplace resummations, as shown in Fig.~\ref{fig:latBorel},
\begin{equation}
  \label{eq:latsum}
    s_{\theta_{\pm}}(\varphi)(z) =
  \frac{1}{z}\int_0^{\re^{\ri(\theta\pm 0)}\infty}\re^{-\zeta/z}\wh{\varphi}(\zeta)
  {\rd }\zeta.
\end{equation}
And in \eqref{eq:disc} we choose $z$ so that
\begin{equation}
  \theta = \arg z = \arg \CA.
\end{equation}
The constant $\ms{s}_{\CA}$ is known as the \emph{Borel residue}.  If
we have a string of singular points $k\CA$ ($k=1,2,\ldots$) along the
ray $\rho_{\arg\CA} = \re^{\ri\arg\CA}\IR_+$, known as the Stokes ray,
the right hand side of \eqref{eq:disc} should be modified to include
contributions from all these non-peturbative saddles
\begin{equation}
  \label{eq:discmulti}
  s_{\theta_+}\varphi(z) - s_{\theta_-}\varphi(z) = \sum_{k=1}^\infty\ms{s}_{k\CA}
  \re^{-k\CA/z} s_{\theta_-}\varphi^{(k\CA)}(z).
\end{equation}

All resurgent series form an algebra, and the analytic formula
\eqref{eq:discmulti} can be represented alternatively as an algebraic
operator in the algebra of resurgent series.  Introducing Stokes
automorphism $\mf{S}_\theta$ associated to the Stokes ray
$\rho_{\theta}$
\begin{equation}
  s_{\theta_+} = s_{\theta_-}\circ\mf{S}_\theta
\end{equation}
so that
\begin{equation}
  \mf{S}_\theta\varphi(z) = \varphi(z) +
  \sum_{k=1}^\infty \ms{s}_{kA}\re^{-k\CA/z}\varphi^{(k\CA)}(z),
\end{equation}
then $\mf{S}_\theta$ is, as its name suggests, an automorphism so that
for two power series $\varphi, \psi$
\begin{equation}
  \mf{S}_\theta(\varphi(z)\psi(z)) = \mf{S}_\theta\varphi(z)
  \mf{S}_\theta \psi(z).
\end{equation}
Another even more powerful way to encode the formula
\eqref{eq:discmulti} is to introduce the alien derivatives
$\dDe{k\CA}$ associated to each Borel singularity related to the
Stokes automorphism $\mf{S}_\theta$ by
\begin{equation}
  \label{eq:S-Del}
  \mf{S}_\theta = \exp\left(\sum_{k=1}^\infty \dDe{k \CA}\right).
\end{equation}
Upon acting on the series $\varphi$, one has
\begin{equation}
  \dDe{k\CA}\varphi(z) = \ms{S}_{k\CA}\varphi^{(k\CA)}(z),
\end{equation}
where the constants $\ms{S}_{k\CA}$ are \emph{Stokes constants}, and
they are combinatoric combinations of $\ms{s}_{k\CA}$.  More
importantly, it can be proved that $\mf{S}_\theta$ are proper
derivations, in the sense that they satisfy the following properties
\begin{itemize}
\item Leibniz rule: if $\varphi,\psi$ are two power series
  \begin{equation}
    \label{eq:leib}
    \dDe{\CA}(\varphi(z)\psi(z)) = \psi(z)\dDe{\CA}\varphi(z)  + \varphi(z)\dDe{\CA}\psi(z).
  \end{equation}
\item Chain rule: if $\psi(x,z)$ is a parametric power series in $z$
  with an auxiliary parameter $x$, and $\psi(z)$ another power series
  in $z$
  \begin{equation}
    \label{eq:chain}
    \dDe{\CA}\varphi(\psi(z),z) = \dDe{\CA}\varphi(x,z)\big|_{x=\psi(z)}
    + \pd_x\varphi(x,z)\dDe{\CA}\psi(z)
  \end{equation}
\item Commutation relation\footnote{Strictly speaking, the commutation
    relation with the expansion parameter $z$ only holds with a
    slightly different convention.  But we will only use the
    commutation relation with auxiliary parameter $x$ in later
    sections.}:
  \begin{equation}
    \label{eq:com}
    [\dDe{\CA},\pd_x] = [\dDe{\CA},\pd_z] = 0.
  \end{equation}
  Thus $\dDe{\CA}$ is a derivation independent of $\pd_x$ and $\pd_z$.
\end{itemize}

\subsection{Resurgent structure of topological string}
\label{sc:top}

Consider topological string with the target space a non-compact
Calabi--Yau threefold $X$.  Let the number of linearly independent
compact 2-cycles and compact 4-cycles be $b_2$ and $b_4$ respectively.
We collect the complexified Kahler moduli of 2-cycles $t^i$
($i=1,\ldots,b_2$) and of 4-cycles $t_{D,\ell}$ ($\ell=1,\ldots,b_4$)
in a vector
\begin{equation}
  \und{\Pi} = \left(
    \begin{array}{c}
      \und{t}_D\\\und{t}\\ 1
    \end{array}\right). 
\end{equation}
where the last entry can be regarded as the trivial Kahler modulus of
a point.  In general $b_2 \geq b_4$, and if $b_2 > b_4$, we can make
the distinction of $b_4$ linear combinations of $t^i$ associated to
2-cycles that intersect with compact 4-cycles, and another $b_2-b_4$
linear combinations of $t^i$ associated to 2-cycles that have zero
intersection numbers with compact 4-cycles.  These linear combinations
are called true moduli and mass parameters, denoted by $t_*^\ell$ and
$t_*^{k+b_4} = m^k$ ($\ell=1,\ldots,b_4$, $k=1,\ldots b_2-b_4$) and we
call the corresponding 2-cycles gauge 2-cycles and flavor 2-cycles, as
they are related to Coulomb moduli and flavor masses in the associated
field theory.  From superstring theory point of view, the Kahler
moduli in the vector $\und{\Pi}$ are also interpreted as the central
charges of D4-, D2-, and D0-branes supported on these cycles.

The moduli space of the Calabi--Yau threefold enjoy special properties
known as the special geometry relations, among which we can define the
prepotential $\CF_0$, a function of $t^i$, so that\footnote{Up to
  normalisation.}
\begin{equation}
  \label{eq:cSpec}
  t_{D,\ell} = \sum_{i=1}^{b_2} C^{i}_{\ell}\frac{\pd \CF_0}{\pd t^i},\quad \ell=1,\ldots,b_4,
\end{equation}
where $C^i_\ell$ are entries of the integer valued $b_2\times b_4$
intersection matrix between compact 2-cycles and compact 4-cycles.
$\CF_0$ is the genus zero component of the free energy, and higher
genus free energies $\CF_g$ can be constructed by coupling the
worldsheet theory to gravity.  Mathematically, the free energies are
defined as the generating function of Gromov-Witten invariants, the
counting of stable holomorphic maps from genus $g$ Riemann surfaces to
2-cycles in the Calabi--Yau $X$, i.e.
\begin{equation}
  \CF_g(\ut) = \sum_{\beta\in H_2(X)} N_{g,\beta} \re^{-t(\beta)},
\end{equation}
where $t(\beta)$ is the complexified volume of the 2-cycle $\beta$.
Collectively, the perturbative free energy of topological string is
\begin{equation}
  \CF^{(0)}_{\text{top}}(\ut,g_s) = \sum_{g\geq 0} \CF_g(\ut) g_s^{2g-2}.
\end{equation}

Through mirror symmetry, the components of the vector $\und{\Pi}$ are
identified with complex structure moduli of the mirror threefold $Y$,
which are periods of the holomorphic $(3,0)$ form over integral
3-cycles in $Y$, or equivalently with periods of the canonical 1-form
over integral 1-cycles in the mirror curve $\Sigma$ that the threefold
$Y$ can reduce to.  Hence $\und{\Pi}$ is also called the period
vector.  The distinction between $t^i$ and $t_{D,\ell}$ corresponds to
a choice of symplectic basis of $H_1(\Sigma)$ consisting of A-cycles
and B-cycles so that the oriented intersection is\footnote{Note that
  in the cases of mirrors to non-compact Calabi--Yau threefolds, it is
  sometimes not possible to make choices of integral basis of cycles
  so that the intersetion matrix is of the form $\delta^{i}_{j}$.}
\begin{equation}
  A^i \cap B_\ell = C^i_\ell.
\end{equation}
And $t^i, t_{D,\ell}$ are correspondingly called the A- and B-periods.
Such a choice is not unique, and a different choice of A- and
B-cycles, known as a different frame, leads to different A- and
B-periods. The frame given by \eqref{eq:cSpec} is called the large
radius frame.  We denote A- and B-periods in a generic frame $\Gamma$
by
\begin{equation}
  \gti, \quad \gtDl,\quad i=1,\ldots,b_2,\quad \ell=1,\ldots,b_4,
\end{equation}
and the special geometry relation \eqref{eq:cSpec} becomes accordingly
\begin{equation}
  \label{eq:cSpecGamma}
  \gtDl = \sum_{i=1}^{b_2} {\gC}^i_\ell
  \frac{\pd \gCF_0}{\pd\gti},\quad \ell=b_4,
\end{equation}
where ${\gC}^i_\ell$ is the intersection matrix of A- and B-cycles in
frame $\Gamma$.  Clearly, the prepotential $\gCF_0$, as well as the
genus $g$ free enegies $\gCF_g$, depend on the choice of frame
$\Gamma$. Although this is not always the case, throughout our paper,
we choose frames so that A- and B-cycles are \emph{integral} cycles,
so that changing a frame amounts to a symplectic transformation in
$Sp(b_3(Y),\IZ)$ of the interal periods.  To find how free energies
change across different frames, it was noted in \cite{Aganagic:2006wq}
that free energies $\gCF_g$ for $g\geq 2$ and $\exp(\gCF_1)$ are
almost holomorphic modular forms, whose modular parameter is
\begin{equation}
  \tau^\Gamma_{\ell m} = \sum_{i,j=1}^{b_2} {\gC}^i_\ell {\gC}^j_m
  \frac{\pd^2\gCF_0}{\pd \gt^i \pd \gt^j},\quad \ell,m=1,\ldots b_4,
\end{equation}
and a change of frame is equivalent to a modular transformation of
these almost holomorphic modular forms.

In the remainder of this section, we will drop the superscript
$\Gamma$ for frame to reduce the notational clutter.  Regardless of
which frame one is at, the genus $g$ free energy $\CF_g(\ut)$ is a
well-defined function, while the perturbative series
$\CF^{(0)}_{\text{top}}(\ut,g_s)$ is a divergent series of Gevrey-1
type \cite{Marino:2006hs,Marino:2007te}
\begin{equation}
  \CF_g(\ut) \sim (2g)!
\end{equation}
and therefore there should be non-perturbative corrections which can
be analysed by resurgence techniques.  It has been found that the
locations of singularities of the Borel transform always coincide with
classical integral periods up to normalisation
\cite{Pasquetti2010,Aniceto2012,Couso-Santamaria2016,Couso-Santamaria2015,Alim:2021mhp,Grassi:2022zuk,Gu:2022sqc}.
More precisely\footnote{See \cite{Gu:2023mgf} for a detailed account
  of the issue of normalisation.},
\begin{equation}
  \label{eq:Almn}
  \CAp =\aleph(\und{p}\cdot\und{t}_D + \und{q}\cdot\und{t} + s),
\end{equation}
where $\aleph = 4\pi^2\ri$, $p^\ell,q_i,s$ are certain integer
numbers.  We will not discuss the cases with $p^\ell = q_i =0$ as they
are trivial.

The alien derivative of the free energy $\CF^{(0)}(\ut,g_s)$ at a
singular point $\CA$ is proportional to the instanton amplitude
associated to this singularity.  In particular, if we have a sequence
of singular points $k\CA$, $k=1,2,\ldots$ along a Stokes ray
$\rho_{\text{arg}\CA}$, and let us denote the instanton amplitude at
the singularity $k\CA$ by $\CF^{(k)}$, the alien derivatives at these
singular points read \cite{Gu:2022sqc,Gu:2023mgf}
\begin{equation}
  \label{eq:DelFTop}
  \dDe{k\CA} \CF^{(0)}_{\text{top}}(\ut,g_s) =
  \frac{\ri}{2\pi} \SSkATop \CF^{(k)}_{\text{top}}(\ut,g_s).
\end{equation}
Here, both the perturbative free energy and the instanton amplitudes
depend on the holomorphic frame of evaluation.  In particular, the
expression of the instanton amplitude depends greatly on the type of
frame.  If a frame, known as an A-frame, is chosen such that
$\CA = \CAp$ is an A-period, i.e.~$p^\ell=0$, then the instanton
amplitude simplifies greatly and we have
\begin{equation}
  \label{eq:FkTop-A}
  \CF^{(k)}_{\text{top},\CA}(\ut,g_s) = \left(\frac{1}{k^2}+\frac{\CA}{k g_s}\right)\re^{-k\CA/g_s},
\end{equation}
where the subscript $\CA$ refers to the A-frame.  If we are not in an
A-frame, the instanton amplitude has more complicated form, but we
will not need them here.

The Stokes constants $\SSkATop$ are very interesting, as it was found
empirically in \cite{Gu:2022sqc,Gu:2023mgf} that they satisfy certain
intriguing properties.  They are integers, and they seem to be frame
independent as well as the same for all the singular points $k\CA$.
Among these properties, the frame independence may be due to the
following reason.  It is known that the holomorphic and frame
dependent free energies $\CF_g(\ut)$ can be lifted to anholomorphic and
frame independent free energies $F_g(\ut,\bar{\ut})$
\cite{Bershadsky:1993cx,Bershadsky:1993ta}, and choosing a frame is
done by sending $\bar{\ut}$ to some fixed value, which can be
interpreted as choosing a different gravitational background
\cite{Witten:1993ed}.  We speculate that \eqref{eq:DelFTop} also holds
when both sides are lifted to anholomorphic amplitudes
\begin{equation}
  \dDe{k\CA} F^{(0)}_{\text{top}}(\ut,\bar{\ut},g_s) =
  \frac{\ri}{2\pi}\SSkATop F^{(k)}_{\text{top}}(\ut,\bar{\ut},g_s).
\end{equation}
The frame independence of the Stokes constants
is then equivalent to the conjecture that they are background
independent.

The purpose of this note is to uncover the nature of the Stokes
constants $\SSkATop$ by relating them to the Stokes constants of the
refined free energy in the Nekrasov--Shatashvili limit.
The perturbative free energy of topological string can be refined to
\begin{equation}
  \CF^{(0)}(\ut,\ep_1,\ep_2) =
  \sum_{n,g\geq 0} (\ep_1+\ep_2)^{2n}(\ep_1\ep_2)^{g-1}\CF_{n,g}(\ut),
\end{equation}
which reduces to the conventional topological string in the limit
$\ep_1  = -\ep_2 = \ri g_s$
\begin{equation}
  \CF^{(0)}_{\text{top}}(\ut,g_s) = \CF^{(0)}(\ut,\ri g_s,-\ri g_s).
\end{equation}
Another interesting limit we can take is the Nekrasov--Shatashvili
limit
\begin{equation}
  \label{eq:FNS}
  \CF^{(0)}_{\text{NS}}(\ut,\hbar) = \lim_{\ep_2\rightarrow 0}
  \ri \ep_2\CF^{(0)}(\ut,\ri\hbar,\ep_2) =: \sum_{n\geq 0} \CF^{\text{NS}}_n(\ut) \hbar^{2n-1}.
\end{equation}
The components $\CF^{\text{NS}}_n(\ut)$ are also almost holomorphic
modular forms and they transform accordingly in a change of frame as
well.

The NS free energies are also Gevrey-1 series in $\hbar$, and we can
similarly perform resurgence analysis.  We find the singularities of
the Borel transform are also located in \eqref{eq:Almn}
\cite{Gu:2022fss}\footnote{In \cite{Gu:2022fss}, locations of Borel
  singularities of NS free energies are found to be
  $4\pi^2(\und{p}\cdot\ut_D + \uq\cdot\ut+s)$.  The
  difference of the factor of $\ri$ is because the authors of
  \cite{Gu:2022fss} used the convention
  $\CF^{(0)}_{\text{NS}} = \sum_n \CF_{(n,0)} \hbar^{2n-1}$ instead of
  the more natural
  $\CF^{(0)}_{\text{NS}} = \sum_n (-1)^n\CF_{(n,0)} \hbar^{2n-1}$ that
  follows from \eqref{eq:FNS}.}.  The alien derivatives of NS free
energy at such a singularity is found to be
\begin{equation}
  \label{eq:DelFWNS}
  \dDe{k\CA} \CF^{(0)}_{\text{NS}}(\ut,\hbar) =
  \frac{\ri\hbar}{2\pi}\SSANS\CF^{(k)}_{\text{NS}}.
\end{equation}
Both perturbative and instantonic free energies are frame dependent.
In the case where $\CA$ is an A-period, i.e.~in an A-frame, one finds
\cite{Gu:2022fss}
\begin{equation}
  \CF^{(\ell)}_{\text{NS},\CA} =
  \frac{(-1)^{k-1}}{k^2}\re^{-k\CA/\hbar}.
  \label{eq:Fl-A}
\end{equation}
In the cases where $\CA$ is not an A-period, i.e.~$\CA$ is given by
\eqref{eq:Almn} with $\und{p}\neq \und{0}$, we can shift the
definition of prepotential $\CF_0$ so that
\begin{equation}
  \label{eq:A-shift}
  \CA = -\frac{1}{2\pi}\sum_{i=1}^{b_2}\sum_{\ell=1}^{b_4}\frac{\pd \CF_0}{\pd t^i} C^i{}_\ell p^\ell
\end{equation}
Then the instanton ampitudes of NS free energy are \cite{Gu:2022fss}
\begin{equation}
  \CF^{(\ell)}_{\text{NS}} =
  \frac{(-1)^{k-1}}{k^2}\re^{-k \CG(\ut,\hbar)/\hbar},
  \label{eq:Fl-B}
\end{equation}
where the quantity $\CG(\ut,\hbar)$ is defined by
\begin{equation}
  \label{eq:CG}
  \CG(\ut,\hbar) = -\frac{\hbar}{2\pi}
  \sum_{i=1}^{b_2}\sum_{\ell=1}^{b_4}
  \frac{\pd \CF^{(0)}_{\text{NS}}(\ut,\hbar)}{\pd t^i}C^i{}_\ell p^\ell.
\end{equation}

Yet again, it was found empirically in \cite{Gu:2022sqc} that the
Stokes constants $\SSANS$ are integers, the same for all $k\CA$, and
seem to be frame independent.  The forms of the alien derivatives
\eqref{eq:DelFWNS} and the properties of the Stokes constants $\SSANS$
have profound consequences.  As pointed out in \cite{Gu:2022fss} (see
also \cite{DelMonte:2022kxh}) and will be reviewed in
Appendix~\ref{sc:qPerds}, they imply, together with the Stokes
transformation properties of Wilson loop amplitudes, that the quantum periods
satisfy the DDP type of formulas for Stokes automorphism
\cite{Dillinger:1993,Delabaere19971:exact,Delabaere1999}, so that the
Stokes constants of quantum periods can be identified with BPS
invariants.  More importantly, the Stokes constants $\SSANS$ are
identified with those of quantum periods \cite{Gu:2022fss}, so that
$\SSANS$ themselves are given by BPS invariants.  More precisely, the
coefficients $(\und{p},\und{q},s)$ in the composition of $\CA$ in
\eqref{eq:Almn} are brane charges.  For instance in the large radius
frame, $(\und{p},\und{q},s)$ are respectively the D4-, D2-, and
D0-brane charges.  $\SSANS$ is then the counting of BPS states of
stable D-brane bound states with brane charges
$\gamma(\CA) = (\und{p},\und{q},s)$; in other words,
\begin{equation}
  \label{eq:NS-BPS}
  \SSANS = \Omega(\gamma(\CA)).
\end{equation}
We will show in Section~\ref{sc:TS-NS} that the Stokes constants
$\SSATop$ of unrefined free energies coincide up to a sign with
$\SSANS$ of NS free energies as in \eqref{eq:TopNSRel}.

\section{Blowup equations of refined topological string}
\label{sc:blowup}

\subsection{Blowup equations in large radius frame}

It was conjecturd \cite{Gu:2017ccq,Huang:2017mis} based on
\cite{Grassi:2016nnt,Sun2016} and checked in many examples that the
blowup equations for supersymmetric gauge theories
\cite{Nakajima:2003pg,Nakajima:2005fg,Nakajima:2009qjc,Gottsche:2006bm}
can be generalised and are satisfied by free energies of topological
string on a local Calabi--Yau threefold $X$.  And it was pointed out
in \cite{Huang:2017mis} that blowup equations can be used to solve
D2-D0 type BPS invariants.  This line of research was later expanded
in
\cite{Gu:2018gmy,Gu:2019dan,Gu:2019pqj,Gu:2020fem,Kim:2019uqw,Kim:2020hhh,Kim:2021gyj,Kim:2023glm,Wang:2023zcb}.
See also related works in
\cite{Nekrasov:2020qcq,Shchechkin:2020ryb,Bershtein:2018zcz,Jeong:2020uxz,Sun:2021lsq}.
The blowup equations will play a crucial role for relating the Stokes
constants of unrefined and NS free energies of topological string.

Let us work in the large radius frame.  The number of linearly
independent compact 2-cycles and 4-cycles in Calabi--Yau threefold $X$
are respectively $b_2$ and $b_4$.  Denote by $\uC$ the $b_2\times b_4$
intersection matrix between compact 2-cycles and 4-cycles.  The
complex Kahler moduli of compact 2-cycles are $\ut$.  Then it was
conjectured that there exist $b_2$-dimensional integer valued vectors
$\und{r}$ satisfying the checkerboard pattern conditions, also known
as flux quantisation conditions, for non-vanishing D2-D0 brane BPS
invariants $N^{\und{d}}_{j_L,j_R}$
\begin{equation}
  2j_L+2j_R +1 \equiv \und{r}\cdot \und{d}\quad \text{mod}\;2,\quad N^{\und{d}}_{j_L,j_R}\neq 0,
\end{equation}
such that the refined free energy of topological string satifies the
so-called blowup equations,
\begin{equation}
  \label{eq:blowup}
  \Lambda(\um,\ep_1,\ep_2) =  \sum_{\und{n}\in\IZ^{b_4}} (-1)^{|\und{n}|}
  \exp\left(\CF_{\text{ref}}(\ut+\frac{\ep_1}{2\pi\ri}\uR,\ep_1,\ep_2-\ep_1)
    +\CF_{\text{ref}}(\ut+\frac{\ep_2}{2\pi\ri}\uR,\ep_1-\ep_2,\ep_2)
    -\CF_{\text{ref}}(\ut,\ep_1,\ep_2)
  \right)
\end{equation}
where $|\und{n}| = n_1+\ldots+n_{b_4}$, and 
\begin{equation}
  \und{R} = \und{C}\cdot\und{n} +\frac{\und{r}}{2}.
\end{equation}
Here the vector $\und{r}$ is in addition subject to the equivalence
relation
\begin{equation}
  \label{eq:equiv}
  \und{r} \sim \und{r} + 2 \und{C}\cdot \und{n}',\quad \und{n}'\in\IZ^{b_4},
\end{equation}
as \eqref{eq:blowup} does not change under this transformation.
Besides the crucial factor $\Lambda(\um,\ep_1,\ep_2)$ depends not on all
the Kahler moduli but only the mass parameters.  We will be interested
in the special cases where $\Lambda$ vanishes identically.  These are
called vanishing blowup equations.

One subtlety concerning the blowup equations as claimed in
\cite{Grassi:2016nnt,Gu:2017ccq,Huang:2017mis} is that the refined
free energies that appear in \eqref{eq:blowup} should be twisted in
the sense that
\begin{equation}
  \CF_{\text{ref}}\rightarrow \wh{\CF}_{\text{ref}}(\ut,\ep_1,\ep_2) = \CF_{\text{ref}}^{\text{pert}}(\ut,\ep_1,\ep_2)
  + \CF_{\text{ref}}^{\text{inst}}(\ut+\frac{\uB}{2},\ep_1,\ep_2),
\end{equation}
where $\uB$, known as the B-field, is a $\IZ_2$ valued
$b_2$-dimensional vector defined by
\begin{equation}
  \uB \equiv \und{r} \; \text{mod} \; (\IZ_2)^{b_2},
\end{equation}
The twisted free energy was introduced so that when a gauge theory
description is available it coincides with the logarithm of the
Nekrasov partition funtion.  Here $F_{\text{ref}}^{\text{inst}}$ is
the instanton contributions, while $F_{\text{ref}}^{\text{pert}}$ is
the sum of classical and 1-loop contributions and it is given
collectively by
\begin{equation}
  \label{eq:Fcls}
  \CF_{\text{ref}}^{\text{pert}}(\ut,\ep_1,\ep_2) =
  \frac{1}{\ep_1\ep_2}
  \left(
    \frac{1}{6}\sum_{i,j,k}a_{ijk}t_it_jt_k  +
    4\pi^2\sum_{i=1}^{s}b_i^{\text{NS}}t_i
  \right) + \sum_{i=1}^s b_i t_i -
  \frac{(\ep_1+\ep_2)^2}{\ep_1\ep_2}\sum_{i=1}^s b_i^{\text{NS}}t_i.
\end{equation}
where $a_{ijk}$ are triple intersection numbers of divisors in $X$ and
$b_i^{\text{NS}}, b_i$ are some other intersection numbers.  The three
terms on the right hand side in \eqref{eq:Fcls} come from
$\CF_{(0,0)}, \CF_{(0,1)}, \CF_{(1,0)}$ respectively.  As it stands,
$\CF_{\text{ref}}^{\text{pert}}$ defined in \eqref{eq:Fcls} does not
have quadratic contributions and it is calculated from the special
geometry relation \eqref{eq:cSpec} using the Frobenius basis of
periods.  If, however, we integrate the special geometry relation
\eqref{eq:cSpec} using the integral basis of periods as we do in this
paper, we would have that
\begin{equation}
  \label{eq:diff-cls}
  \CF_{\text{ref}}^{\text{pert,Frob}}(\ut-\frac{\uB}{2},\ep_1,\ep_2)  =
  \CF_{\text{ref}}^{\text{pert,int'l}}(\ut,\ep_1,\ep_2) +
  \frac{1}{\ep_1\ep_2}\CO(\ut)+\CO(1)+\frac{(\ep_1+\ep_2)^2}{\ep_1\ep_2}\CO(1),
\end{equation}
with an appropriate representation of $B$.  As we will later see in
\eqref{eq:blowup-exp}, the blowup equations only depend on
$\CF_{(0,0)}, \CF_{(0,1)}, \CF_{(1,0)}$ through
\begin{equation}
  \pd_t^{n\geq 2}\CF_{(0,0)}, \; \pd_t^{n\geq 1}\CF_{(0,1)}, \; \pd_t^{n\geq 1}\CF_{(1,0)}
\end{equation}
so that the difference in \eqref{eq:diff-cls} is irrelevant.
In light of this relation, we can use the blowup equations
\eqref{eq:blowup} with the understanding that we can use refined free
energies of topological string based on an integral basis of periods
for the moduli without twist after making the shift
\begin{equation}
  \ut\rightarrow \ut-\uB/2.
\end{equation}

Let us illustrate \eqref{eq:diff-cls} with the simple example of the
local $\IP^2$ model.  This model has a one dimensional moduli space
$\CM$ parametrised by a global modulus $z$.  In the large radius
frame, the integral periods are
\cite{Aganagic:2006wq,Haghighat:2008gw}
\begin{equation}
  \label{eq:IntPiP2}
  \und{\Pi} =
  \begin{pmatrix}
    t_D \\ t \\ 1
  \end{pmatrix} =
  \begin{pmatrix}
    1 & -\frac{1}{2} & \frac{1}{4} \\
    0 & 1 & 0\\
    0 & 0 & 1
  \end{pmatrix} \cdot\und{\Pi}_0
\end{equation}
where $\und{\Pi}_0$ is the Frobenius basis given by
\begin{equation}
  \label{eq:FrobPiP2}
  \und{\Pi}_0 =
  \begin{pmatrix}
    X^{(1,1)} \\ X^{(1)} \\ 1
  \end{pmatrix} =
  \begin{pmatrix}
    \frac{1}{2(2\pi\ri)^2}\left(\log(z)^2 + 2\sigma_1(z)\log(z) +
      \sigma_2(z)\right)\\
    \frac{1}{2\pi\ri}\left(\log(z) + \sigma_1(z)\right)\\
    1
  \end{pmatrix}
\end{equation}
where
\begin{subequations}
  \begin{align}
    \sigma_1(z) = &\sum_{j\geq 1} 3\frac{(3j-1)!}{(j!)^3}(-z)^j,\\
    \sigma_2(z) = &\sum_{j\geq 1} \frac{18}{j!}\frac{(3j-1)!}{(j!)^3}(-z)^j\left(\psi(3j)-\psi(j+1)\right),
  \end{align}
\end{subequations}
with $\psi$ being digamma function.  The special geoemtry relation is
\cite{Aganagic:2006wq}
\begin{equation}
  t_D = -3 (2\pi\ri)^{-3}\pd_t \CF_0(t).
\end{equation}
The prepotential obtained by integrating the special geometry relation
using the integral periods $t_D$ and $t$ is
\begin{equation}
  \CF_0^{\text{int'l}}(t) = -\frac{t^3}{18} + \frac{t^2}{12} -
  \frac{t}{12} + \CO(1)
\end{equation}
while the prepotential obtained by replacing $t_D,t$ with the
Frobenius periods $X^{(1,1)},X^{(1)}$, as practised in
\cite{Huang:2017mis}, is
\begin{equation}
  \CF_0^{\text{Frob}}(t) = -\frac{t^3}{18} + \CO(1)
\end{equation}
and they satisfy \eqref{eq:diff-cls} after taking into account that we
can take $B=1$ in local $\IP^2$ \cite{Huang:2017mis}.

\subsection{Blowup equations in a generic integral frame}

The blowup equations \eqref{eq:blowup} are formulated for free
energies in the large radius frame.  Nevertheless, it is possible to
change the frame and write down the blowup equations in other integral
frames as well.  One way of doing this is using the anholomorphic
blowup equations proposed in \cite{Sun:2021lsq} and choosing the
appropriate holomorphic limit.  Another way is expand the blowup
equations in terms of $\ep_1+\ep_2$ and $\ep_1\ep_2$
\begin{align}
  \re^{\CF_{(0,1)}-\CF_{(1,0)}}
  &\sum_{\un\in\IZ^{b_4}}(-1)^{|\un|}\re^{-\frac{1}{2}R^2\CF''_{(0,0)}}
    (1+(\ep_1+\ep_2)(R\CF'_{(0,1)}+R\CF'_{(1,0)}-\frac{1}{6}R^3\CF'''_{(0,0)}+\ldots)\nn
    =
  &\Lambda_{(0,0)}(\um) + (\ep_1+\ep_2)\Lambda_{(1,0)}(\um)+\ldots.
    \label{eq:blowup-exp}
\end{align}
Here we use the notation
\begin{equation}
  R^k \CF^{(k)}_{(n,g)} = \sum_{i_1,\ldots,i_k}(2\pi\ri)^{-k}R_{i_1}\ldots R_{i_k}\pd_{t_{i_1}}\ldots
  \pd_{t_{i_k}} \CF_{(n,g)}(\ut)
\end{equation}
and $\Lambda_{(n,g)}$ are components of $\Lambda(\um,\ep_1,\ep_2)$
through the expansion
\begin{align}
  \Lambda(\um,\ep_1,\ep_2) = \sum_{n,g\geq 0}\Lambda_{(n,g)}(\ep_1+\ep_2)^{n}(\ep_1\ep_2)^{g}
\end{align}
At each order of  $\ep_1+\ep_2$ and $\ep_1\ep_2$, the left hand side
is a linear sum of
\begin{equation}
  \sum_{\un\in\IZ^{b_4}} (-1)^{|\un|} R^k \re^{-\frac{1}{2}R^2 \CF''_{(0,0)}}
\end{equation}
which are theta constants with modulus $\tau \propto \CF''_{(0,0)}$
and its higher dimensional generalisations.  The coefficients of the
linear sum are products of $\re^{\CF_{(0,1)}-\CF_{(1,0)}}$ and
$\CF^{(k)}_{(n,g)}$ which are almost holomorphic modular forms of
$\tau$.  The identity \eqref{eq:blowup-exp} at each order of
$\ep_1+\ep_2$ and $\ep_1\ep_2$ expansion is an equation of almost
holomorphic modular forms, and they have been checked for various
examples in \cite{Huang:2017mis}.  A frame transformation is then akin
to a modular transformation at each order of \eqref{eq:blowup-exp} and
they can be reassembled into the blowup equation in the corresponding
new frame.

The blowup equation in an arbitrary integral frame takes a form
similar to \eqref{eq:blowup},
\begin{align}
  \gLambda(\um,\ep_1,\ep_2) =  
  &\sum_{\und{n}\in\IZ^{b_4}} (-1)^{|\und{n}|}
    \exp\Big(\gCF_{\text{ref}}(\ugt+\frac{\ep_1}{2\pi\ri}\ugR,\ep_1,\ep_2-\ep_1)
    +\gCF_{\text{ref}}(\ugt+\frac{\ep_2}{2\pi\ri}\ugR,\ep_1-\ep_2,\ep_2)
    \nn
  &-\gCF_{\text{ref}}(\ugt,\ep_1,\ep_2)
    \Big),
  \label{eq:blowup-Gamma}
\end{align}
with
\begin{equation}
  \label{eq:RGamma}
  \ugR = \ugCC\cdot \und{n} + \ugr/2.
\end{equation}
The ingredient $\ugR$ including its coefficients $\ugCC$ and $\ugr$ as
well as $\gLambda(\um,\ep_1,\ep_2)$ may change over different frames.
We will only be interested in vanishing blowup equations so the change
of $\gLambda(\um,\ep_1,\ep_2)$ is trivial as it stays zero across all
frames.

On the other hand, $\ugCC$ and $\ugr$ in $\ugR$ should change
appropriately so that each component in the expansion
\eqref{eq:blowup-exp} transform consistently under modular
transformations.  The properties of $\ugCC$ and $\ugr$ would be
crucial in later sections.  We will only consider frames defined by
integral basis of periods, and in these cases we argue that we always
have
\begin{equation}
  \label{eq:intC}
  \ugCC
  \text{ is an \emph{integer} valued } b_2\times b_4 \text{ matrix.}
\end{equation}
Indeed the sum over $\un\in \IZ^{b_4}$ is a summation over discrete
magnetic flux over the exceptional divisor $\IP^2$ in the spacetime
$\IC^2\cong\IR^4$ blown up at the origin $B\IC^2$ in the field theory
description
\cite{Nakajima:2003pg,Nakajima:2005fg,Gottsche:2006bm,Nakajima:2009qjc},
and each component of the flux vector is associated to an irreducible
compact 4-cycle in the Calabi--Yau $X$
\cite{Grassi:2016nnt,Gu:2017ccq,Huang:2017mis,Kim:2020hhh}.
Therefore, in the case of large radius frame where the moduli
$\gt^i = t^i$ are associated with integral 2-cycles, $\ugCC = \uC$ is
defined as the integer valued intersection matrix of 2-cycles and
4-cycles.  In a generic integral frame, each modulus $\gt^i$ is
associated with either an integral 2-cycle or an integral 4-cycle.  In
the former case, the corresponding row of $\ugCC$ is the integer
valued intersection numbers with 4-cycles; in the latter case, the
corresponding row of $\ugCC$ should be the integer valued
decomposition coefficients in terms of a basis of integral and
irreducible 4-cycles.  We emphysize that in a generic frame, $\ugCC$
is \emph{not} identified with the intersection matrix $\ugC$ given in
\eqref{eq:cSpecGamma}.

Similar to the large radius case, the vector $\ugr$ is defined up to
the equivalence relation
\begin{equation}
  \label{eq:equivGamma}
  \ugr \sim \ugr + 2\ugCC\cdot \und{n}',\quad \und{n}'\in\IZ^{b_4}.
\end{equation}
We also comment that even though we do not have a physics argument,
the vector $\ugr$ also seems to be integer valued in an arbitrary
frame defined by integral periods.  We demonstrate the integrality of
both $\ugCC$ and $\ugr$ through two examples below.

\subsubsection{Local $\IP^2$}
\label{sc:P2}

We first consider the simple example of the local $\IP^2$ model.  The
first two orders of the expansion of the vanishing blowup equations
\eqref{eq:blowup-Gamma} with $\gLambda = 0$ in terms of $\ep_1+\ep_2$
and $\ep_1\ep_2$, similar to \eqref{eq:blowup-exp},
are
\begin{subequations}
  \begin{align}
    \gTheta_0(\gt) = 0 \label{eq:eq1blowupP2}\\
    \gTheta_1(\gt)\left(\gCF'_{(0,1)}(\gt)+\gCF'_{(1,0)}(\gt)\right)
    - \frac{1}{6}\gTheta_3(\gt)
    \gCF'''_{(0,,0)}(\gt) = 0,\label{eq:eq2blowupP2}
  \end{align}
\end{subequations}
where $\gTheta_k(\gt)$ are the theta constants
\begin{equation}
  \label{eq:Thk}
  \gTheta_k(\gt) = \sum_{n\in\IZ} (-1)^n (2\pi\ri)^{-k}
  (\gR)^k \re^{-\frac{1}{2}(\gR)^2\gCF''_{(0,0)}}.
\end{equation}

Consider first the large radius frame, where we will drop all the
superscript $\Gamma$.  These two equations have been checked in
\cite{Huang:2017mis}.  Indeed, we have
\begin{equation}
  R = 3(n+1/2)
\end{equation}
and it is easy to see that \eqref{eq:eq1blowupP2} is satisfied as the
summand of $\Theta_0$ is an odd function of $n$.  Furthermore, let us
introduce the theta constants relevant for the local $\IP^2$ model
\begin{equation}
  a(\tau) = \theta^3\left[
    \begin{array}{c}
      \frac{1}{6} \\ \frac{1}{6}
    \end{array}
  \right](0,\tau),\;
  b(\tau) = \theta^3\left[
    \begin{array}{c}
      \frac{1}{6} \\ \frac{1}{2}
    \end{array}
  \right](0,\tau),\;
  c(\tau) = \theta^3\left[
    \begin{array}{c}
      \frac{1}{6} \\ \frac{5}{6}
    \end{array}
  \right](0,\tau),\;
  d(\tau) = \theta^3\left[
    \begin{array}{c}
      \frac{1}{2} \\ \frac{1}{6}
    \end{array}
  \right](0,\tau),
\end{equation}
with
\begin{equation}
  \theta\left[\ep,\ep'\right](z,\tau) = \sum_{n\in\IZ}\exp(\pi\ri\tau(n+\ep)^2+2\pi\ri(n+\ep)(z+\ep'))
\end{equation}
They have modular weight $3/2$ and enjoy the properties
\begin{equation}
  a(-1/\tau) = \kappa_1\tau^{3/2}c(\tau),\quad b(-1/\tau) = \kappa_2\tau^{3/2}d(\tau),
\end{equation}
where $\kappa_{1,2}$ are roots of unity.  Then the free energies
$\CF_{(0,1)}(t), \CF_{(1,0)}(t)$ are
\cite{Haghighat:2008gw,Huang:2010kf}
\begin{equation}
  \label{eq:F1-MUM}
  \CF_{(0,1)}(t) = -\frac{1}{6}\log(d(\tau)\eta^3(\tau)),\quad
  \CF_{(1,0)}(t) = \frac{1}{6}\log(\eta^3(\tau)/d(\tau)),
\end{equation}
where the modular parameter is
\begin{equation}
  \tau = \frac{\pd t_D}{\pd t} = -3(2\pi\ri)^{-3}\pd^2_t \CF_0.
\end{equation}
Note that here $\CF_{(0,1)}(t)$ is the holomorphic limit of the
anholomorphic
\begin{equation}
  F_{(0,1)}(t,\bar{t}) = -\frac{1}{2}\log \tau_2 \eta^2\bar{\eta}^2 +
  \frac{1}{2}\log \eta/d^{1/3},
\end{equation}
with $\tau_2 = \imag \tau$.  Using expressions of
$\CF_{(0,1)}, \CF_{(1,0)}$ in \eqref{eq:F1-MUM},
\eqref{eq:eq2blowupP2} can be integrated to
\begin{equation}
  \sum_{n\in\IZ}(-1)^n(n+\frac{1}{2})\re^{3\pi\ri(n+\frac{1}{2})^2\tau}
  = \text{Const.} d(\tau)
\end{equation}
which can be checked to high degrees of $q = \exp(2\pi\ri\tau)$
expansion.

As mentioned before, the local $\IP^2$ model has a one dimensional
moduli space $\CM(K_{\IP^2})$ parametrised by a global parameter $z$.
The moduli space of the local $\IP^2$ model has a conifold singularity
at $z = -\frac{1}{27}$, at which the period $t_D$ vanishes.  It is
appropriate then to adopt the conifold frame where $t_D$ is chosen as
the A-period when we are close to the conifold frame, and $t$ as the
B-period.  In the conifold frame, the special geometry relation is
\begin{equation}
  t(t_D) = 3 (2\pi\ri)^{-3} \pd_{t_D}\CF^c_0(t_D).
\end{equation}
The modular parameter is
\begin{equation}
  \tau^c = -\frac{\pd t(t_D)}{\pd t_D} = -3 (2\pi\ri)^{-3}
  \pd^2_{t_D}\CF^c_0(t_D) = -1/\tau,
\end{equation}
so that the first few free energies written as almost holomorphic
modular forms are (up to a constant term)
\begin{equation}
  \label{eq:Fc10}
  \CF^c_{(1,0)}(t_D)= \frac{1}{6}\log(\eta^3(\tau^c)/b(\tau^c)),
\end{equation}
and
\begin{equation}
  F^c_{(0,1)}(t_D,\bar{t}_D) =-\frac{1}{2}\log
  \tau^c_2\eta(\tau^c)^2\bar{\eta}^2(\tau^c)
  +\frac{1}{2}\log \eta(\tau^c)/b^{1/3}(\tau^c),
\end{equation}
whose holomorphic limit is
\begin{equation}
  \label{eq:Fc01}
  \CF^c_{(0,1)}(t_D) = -\frac{1}{6}\log(b(\tau^c)\eta^3(\tau^c)).
\end{equation}
Now \eqref{eq:eq1blowupP2} only holds if
\begin{equation}
  \label{eq:RcP2}
  R^c= \CC^c(n+\frac{1}{2}),
\end{equation}
up to the equivalence relation \eqref{eq:equivGamma} for $r^c$.  Using
\eqref{eq:Fc10},\eqref{eq:Fc01}, the identity \eqref{eq:eq1blowupP2}
can also be integrated to
\begin{equation}
  \sum_{n\in\IZ}(-1)^n(n+\frac{1}{2}) \re^{\frac{(\CC^c)^2}{3} \pi\ri
    (n+\frac{1}{2})^2\tau^c} = \text{Const.}\, b(\tau^c)
\end{equation}
which is only valid for $\CC^c = 1$.  This is consistent with our
prediction for $\CC^c$ as the A-period $t_D$ in the conifold frame is
associated to the irreducible compact 4-cycle.  Together with
\eqref{eq:RcP2}, we can collect the following facts of integrality in
the conifold frame for local $\IP^2$,
\begin{equation}
  \CC^c= 1,\quad r^c = 1.
\end{equation}

\subsubsection{Local $\IP^1\times\IP^1$}
\label{sc:F0}

Here we consider another example of the local $\IP^1\times\IP^1$
model.  This model has one gauge modulus and one mass parameter.  We
restrict ourselves to the case of trivial mass parameter,
corresponding to constraining the two $\IP^1$'s to have the same
complexified Kahler modulus $t$.  In this case, the model also has a
one dimensional moduli space $\CM(K_{\IP^1\times\IP^1})$ parametrised
by a global parameter $z$.

Let us study the vanishing blowup equations.  The first two equations
from expanding the vanishing blowup equations in terms of
$\ep_1+\ep_2$ and $\ep_1\ep_2$ are still
\eqref{eq:eq1blowupP2},\eqref{eq:eq2blowupP2}.  We consider again the
large radius frame first.  In the massless local $\IP^1\times\IP^1$
model, we have \cite{Huang:2017mis}
\begin{equation}
  R = 2(n+1/2),
\end{equation}
and the free energies $\CF_{(0,1)}(t),\CF_{(1,0)}(t)$ are respectively
\cite{Haghighat:2008gw,Huang:2010kf}
\begin{equation}
  \CF_{(0,1)}(t) = -\log\eta(\tau),\quad
  \CF_{(1,0)}(t) = -\frac{1}{6}\log\theta_2^2(\tau)/(\theta_3(\tau)\theta_4(\tau)),
\end{equation}
where the modular parameter is
\begin{equation}
  \tau = \frac{\pd t_D(t)}{\pd t} = -2(2\pi\ri)^{-3}\pd_t^2\CF_0.
\end{equation}
Here we normalise the periods similar to \eqref{eq:FrobPiP2} in local
$\IP^2$, namely
\begin{equation}
  \label{eq:norm}
  t = X^{(1)} = \frac{1}{2\pi\ri}\log z + \ldots,\qquad
  t_D = X^{(1,1)}+ \ldots = \frac{1}{2(2\pi\ri)^2}\log^2z + \ldots.
\end{equation}
Furthermore, as in local $\IP^2$, $\CF_{(0,1)}(t)$ is the holomorphic
limit of an anholomorphic free energy which reads
\begin{equation}
  F_{(0,1)}(t,\bar{t}) = -\frac{1}{2}\log\tau_2\eta^2\bar{\eta}^2.
\end{equation}
We can then check that \eqref{eq:eq1blowupP2} naturally holds as the
summand of $\Theta_0$ is yet again an odd function of $n$, while
\eqref{eq:eq2blowupP2} can be integrated to
\begin{equation}
  \sum_{n\in\IZ} (-1)^n
  (n+\frac{1}{2})\re^{2\pi\ri(n+\frac{1}{2})^2\tau} = \text{Const.} \frac{\eta^3(\tau)\theta_2(\tau)}{\sqrt{\theta_3(\tau)\theta_4(\tau)}}
\end{equation}
which can be checked to high degrees of $q = \exp(2\pi\ri\tau)$
expansion.

The moduli space $\CM(K_{\IP^1\times\IP^1})$ of the massless local
$\IP^1\times\IP^1$ model has a conifold singularity at
$z = \frac{1}{16}$, at which the period $t_D$ vanishes.  It is then
suitable to choose the conifold frame where $t_D$ is selected as the
A-period when we are close to the conifold point, and $t$ as the
B-period.  In the conifold frame, the special geoemtry relation is
\begin{equation}
  t = 2(2\pi\ri)^{-3} \pd_{t_D}\CF_0^c(t_D)
\end{equation}
and the modular parameter is
\begin{equation}
  \tau^c = -\frac{\pd t}{\pd t_D} =
  -2(2\pi\ri)^{-3}\pd_{t_D}^2\CF_0^c(t_D) = -\frac{1}{\tau}.
\end{equation}
Using that
\begin{equation}
  \theta_2(-1/\tau) = \kappa_1' \tau^{1/2}\theta_4(\tau),\quad
  \theta_3(-1/\tau) = \kappa_2' \tau^{1/2}\theta_3(\tau),
\end{equation}
where $\kappa'_{1,2}$ are roots of unity, it is easy to find the first
few free energies are (up to a constant term)
\begin{equation}
  \label{eq:Fc10F0}
  \CF_{(1,0)}^c(t_D) = -\frac{1}{6}\log\frac{\theta_4^2(\tau^c)}{\theta_3(\tau^c)\theta_2(\tau^c)}
\end{equation}
and
\begin{equation}
  F^c_{(0,1)}(t_D,\bar{t}_D) = -\frac{1}{2}\log\tau_c^2\eta^2(\tau^c)\bar{\eta}^2(\bar{\tau}^c)
\end{equation}
whose holomorphic limit is
\begin{equation}
  \label{eq:Fc01F0}
  \CF^c_{(0,1)}(t_D) = -\log(\eta(\tau^c)).
\end{equation}
We find then again that \eqref{eq:eq1blowupP2}  holds if and only if
\begin{equation}
  \label{eq:RcF0}
  R^c = \CC^c(n+\frac{1}{2})
\end{equation}
up to the equivalence relation \eqref{eq:equivGamma} for $r^c$.  Using
\eqref{eq:Fc10F0},\eqref{eq:Fc01F0}, we find that
\eqref{eq:eq2blowupP2} can also be integrated to
\begin{equation}
  \sum_{n\in\IZ}(-1)^n
  (n+\frac{1}{2})\re^{\frac{(\CC^c)^2}{2}\pi\ri(n+\frac{1}{2})^2\tau^c}
  \text{Const.} \frac{\eta^3(\tau^c)\theta_4(\tau^c)}{\sqrt{\theta_2(\tau^c)\theta_3(\tau^c)}}.
\end{equation}
This is only valid for $\CC^c=1$, which is consistent with our
prediction for $\CC^c$ as the A-period $t_D$ in the conifold frame is
associated to the irreducible compact 4-cycle.  Together with
\eqref{eq:RcF0} we collect the following facts of integrality in the
conifold frame for massless local $\IP^1\times \IP^1$,
\begin{equation}
  \CC^c = 1,\quad r^c = 1.
\end{equation}

\section{Relation between unrefined and NS Stokes constants}
\label{sc:TS-NS}

In this section, we reveal the intimate connection between the Stokes
constants of unrefined and NS free energies.
The starting point is the observation in
\cite{Grassi:2016nnt,Huang:2017mis} that in the limit $\ep_2 = 0$ the
blowup equations of \eqref{eq:blowup} of the vanishing type in the
large radius frame become the so-called compactibility formulas which
relate the unrefined and the NS free energies of topological string.
The same is true for vanishing blowup equations in an arbitrary
integral frame, and the corresponding compactibility formula reads
\begin{equation}
  \label{eq:compt}
  0 = \sum_{\un\in\IZ^{b_4}} (-1)^{|\un|}
  \exp\left(\CF_{\text{top}}(\ut+\frac{\hbar}{2\pi}\uR,\hbar)-\ri\sum_{i=1}^{b_2}\sum_{\ell=1}^{b_4}
    \CC^i{}_{\ell}n_\ell\frac{\pd}{2\pi\ri \pd t^i}
    \CF_{\text{NS}}(\ut,\hbar)\right),
\end{equation}
where $\uR$ is given in \eqref{eq:RGamma}.  In this section, we will
drop the superscript $\Gamma$ indicative of frame to reduce notational
clutter.

Let $\CA$ be a point of the type \eqref{eq:Almn} in the Borel plane.
We take the compactibility formula \eqref{eq:compt} in an A-frame
where $\CA$ is a classical A-period so that $\und{p} = \und{0}$, and
apply the alien derivative $\dDe{k\CA}$.  After using the Leibniz rule
\eqref{eq:leib}, we find that
\begin{align}
  0 =
  &\sum_{\un\in\IZ^{b_4}} (-1)^{|\un|}
    \exp\left(\CF_{\text{top}}(\ut+\frac{\hbar}{2\pi}\uR,\hbar)-\ri\sum_{i=1}^{b_2}\sum_{\ell=1}^{b_4}
    \CC^i{}_{\ell}n_\ell\frac{\pd}{2\pi\ri\pd t^i}\CF_{\text{NS}}(\ut,\hbar)\right)\nn
  &\times\left(\dDe{k\CA}\CF_{\text{top}}(\ut+\frac{\hbar}{2\pi}\uR,\hbar) -\ri\sum_{i=1}^{b_2}\sum_{\ell=1}^{b_4}
    \CC^i{}_{\ell}n_\ell\frac{\pd}{2\pi\ri\pd t^i}\dDe{k\CA}\CF_{\text{NS}}(\ut,\hbar)
    \right).
    \label{eq:DelCompt}
\end{align}
Since both the unrefined and NS free energies are evaluated in an
A-frame, their alien derivatives are simple, given by
\eqref{eq:DelFTop},\eqref{eq:FkTop-A},\eqref{eq:DelFWNS},\eqref{eq:Fl-A}.
We then use the commutation relation \eqref{eq:com} to find
\begin{equation}
  \frac{\pd}{2\pi\ri\pd t^i} \dDe{k\CA}\CF_{\text{NS}}(\ut,\hbar) = \ri
  q_i \SSANS \frac{(-1)^k}{k}\re^{-k\CA/\hbar}
\end{equation}
and use the chain rule \eqref{eq:chain} to find
\begin{equation}
  \dDe{k\CA}\CF_{\text{top}}(\ut+\frac{\hbar}{2\pi}\uR,\hbar) =
  \frac{\ri}{2\pi}\SSkATop
  \left(\frac{1}{k^2}+\frac{\CA}{k\hbar} + \frac{\pi\ri \uq\cdot \ur}{k} +
    \frac{2\pi\ri}{k}\uq\cdot \uCC\cdot \un\right)(-1)^{k\uq\cdot \ur}\re^{-k \CA/\hbar}
\end{equation}
where we have used that $\uq\cdot\uCC\cdot\un \in \IZ$ due to the
integrality condition \eqref{eq:intC}.  We plug these equations in the
second line of \eqref{eq:DelCompt} and drop any component which
vanishes due to the compactibility formula, and we arrive at the
crucial equation
\begin{align}
  0 =
  &\left(-\SSkATop(-1)^{k (\uq\cdot \ur)} + \SSANS (-1)^k
    \right)\frac{1}{k}\re^{-k\CA/\hbar}\nn
  &\times\sum_{\un\in\IZ^{b_4}} \uq\cdot \uCC\cdot \un (-1)^{|\un|}
    \exp\left(\CF_{\text{top}}(\ut+\frac{\hbar}{2\pi}\uR,\hbar)-\ri\sum_{i=1}^{b_2}\sum_{\ell=1}^{b_4}
    \CC^i{}_{\ell}n_\ell\frac{\pd}{2\pi\ri\pd t^i}\CF_{\text{NS}}(\ut,\hbar)\right).
    \label{eq:SS}
\end{align}

We make the distinction between two cases.
If the singularity $\CA$ is such that
\begin{equation}
  \uq\cdot \uCC = \und{0},
  \label{eq:qC=0}
\end{equation}
the second line vanishes and \eqref{eq:SS} gives no constraint between
the two types of Stokes constants.  Geometrically, condition
\eqref{eq:qC=0} corresponds to flavor 2-cycles in Calabi--Yau
threefold $X$, which are 2-cycles that have zero intersection numbers
with compact 4-cycles.  Flavor 2-cycles are not expected to support
BPS states.

If, on the other hand, the condition \eqref{eq:qC=0} is not satisfied,
we argue that one can at least find one vector $\ur$ for vanishing
blowup equations so that the second line of \eqref{eq:SS} does not
vanish.  Following \eqref{eq:blowup-exp}, the leading term in the
$\hbar$ expansion of the second line of \eqref{eq:SS} is
\begin{equation}
  \label{eq:SSleading}
  \sum_{\un \in \IZ^{b_4}} (\uq\cdot \uCC\cdot \un)(-1)^{|\un|} \re^{-\frac{1}{2}R^2\CF''_{(0,0)}}.
\end{equation}
Recall from \eqref{eq:eq1blowupP2} that the leading term in the
vanishing blowup equations is
\begin{equation}
  \label{eq:vbuleading}
  \sum_{\un \in \IZ^{b_4}} (-1)^{|\un|}
  \re^{-\frac{1}{2}R^2\CF''_{(0,0)}} = 0,
\end{equation}
and it has been conjetured in \cite{Huang:2017mis} that in the large
radius frame, once $\uCC$ is known, any integer valued vector $\ur$
with which \eqref{eq:vbuleading} holds is a suitable $\ur$ vector for
vanishing blowup equations \eqref{eq:blowup} with $\Lambda = 0$.
Among these suitable $\ur$ vectors, a special one is the one that
makes the summand of \eqref{eq:vbuleading} an odd function of $\un$.
We denote such an $\ur$ vector by $\ur_{\text{odd}}$, and we
conjecture that it exists in any integral frame, as the structure of
vanishing blowup equations are similar across different frames.  In
the local $\IP^2$ model discussed in Section~\ref{sc:P2},
$\ur_{\text{odd}} = 3$ in the large radius frame, and
$\ur_{\text{odd}}= 1$ in the conifold frame.  Similarly, in the local
$\IP^1\times\IP^1$ model discussed in Section~\ref{sc:F0},
$\ur_{\text{odd}} = 2$ in the large radius frame, and
$\ur_{\text{odd}}= 1$ in the conifold frame.  Such an
$\ur_{\text{odd}}$ can also be found in all the examples discussed in
\cite{Huang:2017mis}.  Now if we take $\ur = \ur_{\text{odd}}$ in
\eqref{eq:SSleading}, it will no longer be zero as the linear term
$\uq\cdot\uCC\cdot\un$ changes the summand from an odd function to an
even function of $\un$.

As the second line of \eqref{eq:SS} does not vanish, we can naturally
conclude that the Stokes constants of unrefined free energy and those
of NS free energy be the same up to a sign
\begin{equation}
  \label{eq:TopNSRel}
  \boxed{\SSkATop = (-1)^{k(\uq\cdot\ur_{\text{odd}}-1)} \SSANS.}
\end{equation}
This is our key formula.  And thanks to \eqref{eq:NS-BPS}, it implies that
\begin{equation}
  \label{eq:SSTopBPS}
  \boxed{\SSkATop \;\fallingdotseq\; \Omega(\gamma(\CA)),}
\end{equation}
i.e.~the Stokes constant of unrefined free energy of topological
string for the Borel singularity $k\CA$, which is not associated with
flavor 2-cycles, coincide with the BPS invariant $\Omega(\gamma(\CA))$
where $\gamma(\CA)$ is the brane charge associated with $\CA$,
possibly up to a sign as indicated by the dotted equality sign
$\fallingdotseq$.

\section{Discussion}
\label{sc:disc}

In this note, we are interested in the relationship between Stokes
constants of unrefined perturbative free energies and those of refined
perturbative free energies in the Nekrasov--Shatashvili limit of
topological string theory on a non-compact Calabi--Yau threefold.  It
was observed in \cite{Gu:2022fss,Gu:2022sqc} with the example of local
$\IP^2$ that both perturbative series have Borel singularities located
at classical integral periods of mirror Calabi--Yau in the B-model,
and the Stokes constants of the two perturbative series at the same
Borel singularity might be related.  We confirm this observation and
demonstrate, using the formulation of the blowup equations in a
generic integral frame taken to certain special limit, that this
should be true on a generic non-compact Calabi--Yau threefold.  More
precisely, as long as the Borel singularity does not correspond to
some 2-cycles that intersect only with non-compact 4-cycles in the
A-model, the Stokes constants of the unrefined and the NS perturbative
free energies must be the same up to a sign.

It was argued in
\cite{Bridgeland:2016nqw,Bridgeland:2017vbr,Alim:2021mhp} that the
Stokes constants of unrefined topological string free energy for
non-compact Calabi--Yau threefold should be \emph{related} to BPS
invariants, although as far as concrete constructions are concerned
only the simplest Calabi--Yau threefold, the resolved conifold, is
considered.
Similar statements for more generic non-compact Calabi--Yau
\cite{Gu:2021ize,Gu:2022sqc} and even for compact Calabi--Yau
threefolds \cite{Gu:2023mgf} have been proposed.
In this paper we give strong support for these statements.  In fact,
since the Stokes constants of NS free energies can be shown
\cite{Gu:2022fss} to coincide with the Stokes constants of quantum
periods, and therefore can be interpreted as BPS invariants, the
results of this paper imply immediately that the Stokes constants of
unrefined free energies on a non-compact Calabi--Yau threefold can
similarly be \emph{identified} as counting of BPS states, i.e.~stable
D4-D2-D0 brane bound states in type IIA string.  Similary conclusions
have been reached in \cite{Iwaki:2023rst}, where a closed-form formula
for Stokes automorphisms of unrefined topological string free energies
has been provided, which resembles the DDP formula of quantum periods.

This paper opens many new directions to explore.  First of all, our
demonstration of the BPS interpretation of Stokes constants of
unrefined free energies hinges on two crucial conjectures, the blowup
equations for refined topological string free energies in a generic
integral frame, and the identification of Stokes constants of NS free
energies as BPS invariants.  More evidences or maybe proofs are needed
for these two conjectures.

Furthermore, the argument in this paper is indirect and is only valid
for non-compact Calabi--Yau threefolds.  A direct argument or even
proof, potentially also valid for compact Calabi--Yau threefold,
possibly along the line of \cite{Iwaki:2023rst}, will be very
desirable.

Finally, the result of this paper suggests using resurgence techniques
to systematically study BPS invariants of Calabi--Yau threefolds in
different loci of moduli space and to study their stability
structures.  It would be interesting to compare with the BPS spectrum
of local $\IF_0$ in \cite{Longhi:2021qvz} and of local $\IP^2$ in
\cite{Bousseau:2022snm} computed with other techniques.

\section*{Acknowledgement}

We would like to thank Alba Grassi, Lotte Hollands, Min-xin Huang,
Albrecht Klemm, Marcos Mari\~no, Boris Pioline, Kaiwen Sun for useful
discussions.  We also thank the hospitality of the Mainz Institute for
Theoretical Physics (MITP) of Johannes Gutenberg University, the host
of the workshop ``Spectral Theory, Algebraic Geometry and Strings''
(STAGS2023), during which part of this work was completed.  J.G.~is
supported by the startup funding No.~4007022316 of the Southeast
University.

\vspace{2ex}

\noindent\emph{Special thanks: During the write-up of this paper,
  Marcos Mari\~no sent us his draft together with Kohei Iwaki on a
  related subject: the derivation of a closed-form formula for Stokes
  automorphisms of unrefined topological string free energies
  \cite{Iwaki:2023rst}.  Although the methods and the concrete results
  of their paper are different from ours, we both give strong support
  to the idea that Stokes constants of unrefined free energies are
  given by BPS invariants.  We thank Kohei and Marcos for agreeing to
  coordinate the submission of these two papers.}

\appendix

\section{Stokes automorphisms of quantum periods}
\label{sc:qPerds}

In this Appendix we recall and slightly generalise the derivation in
\cite{Gu:2022fss} that the Stokes automorphisms of quantum periods in
topological string should follow the DDP formulas
\cite{Dillinger:1993,Delabaere19971:exact,Delabaere1999}, and that the
Stokes constants of NS free energies coincide with those of quantum
periods, which combined together imply that the Stokes constants of NS
free energies of topological string can be identified with BPS
invariants.  We will present the derivation in the large radius frame,
but the derivation in a generic integral frame is similar.

Let us first expand on the definition of gauge moduli $t_*^\ell$ and
flavor masses $t_*^{k+b_4} = m^k$ for $\ell=1,\ldots,b_4$ and
$k=1,\ldots,b_2-b_4$ introduced in Section~\ref{sc:top}.  One possible
way of choosing them is to complete the $b_2\times b_4$ intersection
matrix $C^i{}_\ell$ to an $b_2\times b_2$ matrix $\hat{C}^i{}_j$ of
full rank by including intersections of 2-cycles with non-compact
4-cycles.  Then the gauge moduli $t_*^\ell$ and the flavor masses
$t_*^{k+b_4} = m^k$ can be chosen as
\begin{equation}
  t^\ell_* = \sum_{j=1}^{b_2}(\hat{C}^{-1})^\ell_j t^j,\quad m^k = \sum_{j=1}^{b_2}(\hat{C}^{-1})^{k+b_4}_jt^j.
\end{equation}
Conversely, we have
\begin{equation}
  t^i = \sum_{\ell=1}^{b_4}C^i_\ell t_*^\ell + \sum_{k=1}^{b_2-b_4}\hat{C}^i_{k+b_4}m^k.
\end{equation}
Then the special geometry relation \eqref{eq:cSpec} can be written as
\begin{equation}
  t_{D,\ell}(\uz)=
  (2\pi\ri)^{-3}
  \frac{\pd\CF_0(\uz(\ut))}{\pd t_*^\ell}\Big|_{\ut_*=\ut_*(\uz)},  \quad
  \ell=1,\ldots,b_4.
  \label{eq:cSpecTrue}
\end{equation}
Here we have included the proper normalisation prefactor consistent
with the normalisation of periods as in \eqref{eq:norm}.  And we
denote by $\uz$ the global moduli of the model, and $\uz(\ut)$ are the
classical mirror map.  The classical periods $t_*^\ell(\uz)$ and
$t_{D,\ell}(\uz)$ (but not the mass parameters $m^k$) can be promoted to
quantum periods $t_*^\ell(\uz,\hbar)$ and $t_{D,\ell}(\uz,\hbar)$, and it
was observed in \cite{Aganagic:2011mi} that they satisfy the quantum
special geometry relation
\begin{equation}
  \label{eq:qSpec}
  t_{D,\ell}(\uz,\hbar) =
  (2\pi\ri)^{-3} 
  \frac{\hbar\pd \CF^{(0)}_{\text{NS}}(\uz(\ut),\hbar)}{\pd t_*^\ell}
  \Big|_{\ut_* = \ut_*(\uz,\hbar)},\quad \ell=1,\ldots,b_4,
\end{equation}
which states that quantum A- and B-periods satisfy the same
relationship as classical A- and B-periods, as long as the global
moduli $\uz$ in the NS free energy are related to the \emph{quantum}
A-periods $t(\uz,\hbar)$ through the \emph{classical} mirror map
$\uz(\ut)$.

Another type of quantities we will need are the NS Wilson loop amplitudes in
different representations $\rho_\ell$ of gauge group, which can be
defined for 4d or 5d gauge theories that topological string engineers
on a non-compact Calabi-Yau threefold\footnote{For the case of local
  $\IP^2$, which has no apparent gauge group, a notion of Wilson loop amplitudes
  can also be defined \cite{Kim:2021gyj}.}, and we refer to
\cite{Kim:2021gyj,Huang:2022hdo,Wang:2023zcb} for background and
references.  They are also frame dependent, and each of them is a
Gevrey-1 power series in $\hbar$
\begin{equation}
  \label{eq:WNS}
  \CW^{(0)}_{\text{NS},\rho_\ell}(\uz,\hbar) =
  \sum_{n\geq 0} \CW^{\text{NS}}_{\rho_\ell,n}(\uz) \hbar^{2n}.
\end{equation}
The Stokes automorphisms of NS Wilson loop amplitudes have been studied in
\cite{Gu:2022fss}.  The crucial fact we will use is that the NS Wilson
loop amplitudes have Borel singularities also at $\CAp$ as defined in
\eqref{eq:Almn}, but only when $\CAp$ is a B-period.  In other words
\begin{equation}
  \label{eq:DelW0}
  \dDe{k\CA} \CW^{(0)}_{\text{NS},\rho_\ell}(\uz,\hbar) = 0
\end{equation}
if $\CA$ is an A-period.

In addition, it was observed in \cite{Gu:2022fss} that, similar to
\eqref{eq:qSpec}, the NS Wilson loop amplitudes offer another set of relations
between quantum periods
\begin{equation}
  \label{eq:qWil}
  \CW^{(0)}_{\text{NS},\rho_\ell}(\uz(\ut),\hbar)\Big|_{\ut_*=\ut_*(\uz,\hbar)} = \CW^{\text{NS}}_{\rho_\ell,0}(\uz),
\end{equation}
in other words, the quantum Wilson loop amplitude as a power series of $\hbar$
reduces to the classical Wilson loop amplitude which does not depend on $\hbar$,
if the global moduli $\uz$ in the quantum Wilson loop amplitude are related to
the quantum A-periods via the classical mirror maps $\uz(\ut)$.

We will then show that using the identities \eqref{eq:qSpec} and
\eqref{eq:qWil} we can deduce the properties of Stokes automorphisms
of quantum periods from those of NS free energies and NS Wilson loop amplitudes.
Let $\CA$ be a Borel singularity of the type $\CAp$ as in
\eqref{eq:Almn}.
By appying the alien derivative $\dDe{k\CA}$ on both sides of
\eqref{eq:qWil} and using the chain rule \eqref{eq:chain}, we find the
relation of alien derivatives
\begin{equation}
  \label{eq:qWilRel}
  0 = \dDe{k\CA}\CW^{(0)}_{\text{NS},\rho_\ell}(\uz(\ut),\hbar)
  \Big|_{\ut_*=\ut_*(\uz,\hbar)}
  + \sum_{n=1}^{b_4}\pd_{t_*^n}\CW^{(0)}_{\text{NS},\rho_\ell}(\uz(\ut),\hbar)
  \Big|_{\ut_*=\ut_*(\uz,\hbar)}
  \dDe{k\CA}t_*^n(\uz,\hbar).
\end{equation}
Similarly, by applying the alien derivative $\dDe{k\CA}$ on both sides
of \eqref{eq:qSpec}, we find
\begin{equation}
  \label{eq:qSpecRel}
  (2\pi\ri)^3\dDe{k\CA}t_{D,\ell}(\uz,\hbar) = \hbar \pd_{t_*^\ell} \dDe{k\CA}
  \CF^{(0)}_{\text{NS}}(\uz(\ut),\hbar)\Big|_{\ut_*=\ut_*(\uz,\hbar)} +
  \hbar\sum_{n=1}^{b_4}\pd_{t_*^\ell}\pd_{t_*^n}
  \CF^{(0)}_{\text{NS}}(\uz(\ut),\hbar)\Big|_{\ut_*=\ut_*(\uz,\hbar)}
  \dDe{k\CA}  t_*^n(\uz,\hbar).
\end{equation}
Note that this equation demonstrates the clear difference between
using the classical mirror map and the quantum mirror map when
computing the Stokes automorphisms or alien derivatives of NS free
energies.  \eqref{eq:qWilRel} and \eqref{eq:qSpecRel} are two crucial
identities we will make heavy use of.

Let us first assume that $\CA$ is an A-period,
i.e.~$\und{p} = \und{0}$.  From \eqref{eq:qWilRel} and
\eqref{eq:DelW0} we find immediately
\begin{equation}
  \label{eq:Dela-A}
  \dDe{k\CA}t_*^\ell(\uz,\hbar) = 0,\quad \ell=1,\ldots,b_4.
\end{equation}
In other words, the quantum A-periods have vanishing Stokes constants.
Next by using \eqref{eq:qSpecRel} and \eqref{eq:Dela-A} as well as 
\eqref{eq:DelFWNS}, \eqref{eq:Fl-A}, we find
\begin{equation}
  \label{eq:DelaD-A}
  \dDe{k\CA} t_{D,\ell}(\uz,\hbar) =
  -(\uq\cdot\uC)_\ell\frac{\hbar}{\aleph} \SSANS \frac{(-1)^{k-1}}{k}
  \re^{-k\CA(\uz(\ut))/\hbar}\Big|_{\ut_*=\ut_*(\uz,\hbar)}.
\end{equation}
\eqref{eq:Dela-A} and \eqref{eq:DelaD-A} imply the Stokes
automorphisms of quantum A- and B-periods across $\rho_{\arg\CA}$ when
$\CA$ is an A-perod are
\begin{subequations}
  \begin{align}
    \mf{S}_{\arg\CA(\uz)} t_*^\ell(\uz,\hbar) =
    & t_*^\ell(\uz,\hbar),
      \label{eq:St-A}\\
    \mf{S}_{\arg\CA(\uz)}t_{D,\ell}(\uz,\hbar) =
    &t_{D,\ell}(\uz,\hbar)-(\uq\cdot\uC)_\ell\frac{\hbar}{\aleph}
      \SSANS\log(1+\re^{-\CA(\uz(\ut))/\hbar})\Big|_{\ut_*=\ut_*(\uz,\hbar)}.
      \label{eq:StD-A}
  \end{align}
\end{subequations}
Note that the exponent of the right hand side of \eqref{eq:StD-A} is
in fact a quantum A-period
\begin{equation}
  \CA(\uz(\ut))\big|_{t_*=t_*(z,\hbar)} = \aleph(\uq\cdot \ut(\uz,\hbar)+s)
\end{equation}
and thus the Stokes automorphism of the quantum B-period
$t_D(\uz,\hbar)$ is expressed in terms of the quantum A-period
associated to $\CA(\uz)$.

Next we consider the case where $\CA$ is no longer
an A-period, i.e.~$\und{p}\neq \und{0}$.
We will again make the shift of prepotential $\CF_0$ as in
\eqref{eq:A-shift}.
If we were in a different frame where $t_{D,\ell}$ were A-periods,
$\CA$ would be an A-period as it is a linear combination of
$t_{D,\ell}$ by virtue of \eqref{eq:A-shift}.  Therefore by symmetry
argument, we should have
\begin{equation}
  \label{eq:DelaD-B}
  \dDe{k\CA}t_{D,\ell}(\uz,\hbar) = 0.
\end{equation}
If we now impose \eqref{eq:DelaD-B} and use \eqref{eq:qSpecRel}
together with \eqref{eq:DelFWNS}, \eqref{eq:Fl-B}, we will find that
\begin{equation}
  \label{eq:Dela-B}
  \dDe{k\CA}t_*^\ell(\uz,\hbar) =
  p^\ell\frac{\hbar}{\aleph} \SSANS \frac{(-1)^{k-1}}{k}
  \re^{-k \CG(\ut,\hbar)/\hbar}\Big|_{\ut_*=\ut_*(\uz,\hbar)}.
\end{equation}
From \eqref{eq:DelaD-B} and \eqref{eq:Dela-B} we find the Stokes
automorphisms of quantum A- and B-periods across $\rho_{\arg\CA}$ when
$\CA$ is a B-period are
\begin{subequations}
  \begin{align}
    \mf{S}_{\arg\CA(\uz)} t_*^\ell(\uz,\hbar) =
    &t_*^\ell(\uz,\hbar)+p^\ell\frac{\hbar}{\aleph}\SSANS
      \log(1+\re^{-\CG(\ut,\hbar)/\hbar})\Big|_{\ut_*=\ut_*(\uz,\hbar)},
    \label{eq:St-B}\\
    \mf{S}_{\arg\CA(\uz)} t_{D,\ell}(\uz,\hbar) =
    &t_{D,\ell}(\uz,\hbar).
      \label{eq:StD-B}
  \end{align}
\end{subequations}
Note that $\CG(\ut,\hbar)$ as defined in \eqref{eq:CG} is the quantum
deformation of the B-period $\CA$, and after replacing $\ut_*$ by the
quantum A-periods,
\begin{equation}
  \CG(\und{t},\hbar)\Big|_{\ut_*=\ut_*(\uz,\hbar)} =  \frac{\hbar}{2\pi}\sum_{i=1}^{b_2}\sum_{\ell=1}^{b_4}\pd_{t^i}
  \CF^{(0)}_{\text{NS}}(\uz(\ut),\hbar)\Big|_{\ut_*=\ut_*(\uz,\hbar)} C^i{}_\ell p^\ell
  = \CA(\uz(\ut))\Big|_{\ut_*=\ut_*(\uz,\hbar)},
\end{equation}
it becomes the corresponding quantum B-period.  Therefore the Stokes
automorphisms of the quantum A-periods $t_*^\ell(\uz,\hbar)$ are
expressed in terms of the quantum B-periods associated to the B-period
$\CA(\uz)$.

By introducing the symplectic pairing $\vev{\gamma,\gamma'}$ of two
quantum periods $\Pi_\gamma(\uz,\hbar), \Pi_{\gamma'}(\uz,\hbar)$
where
\begin{equation}
  \gamma= (\und{p},\und{q},s),\quad \gamma' = (\und{p}',\und{q}',s')
\end{equation}
and
\begin{equation}
  \vev{\gamma,\gamma'} = \und{q}'\cdot\und{C}\cdot\und{p} -\und{q}\cdot\und{C}\cdot\und{p}',
\end{equation}
as well as the Voros symbol
\begin{equation}
  \CV_\gamma(\uz,\hbar)  = \exp(-\aleph \Pi_\gamma(\uz,\hbar)/\hbar)
\end{equation}
the Stokes automorphisms
\eqref{eq:St-A},\eqref{eq:StD-A},\eqref{eq:St-B},\eqref{eq:StD-B} can
be summarised succinctly by the DDP type of formulas
\cite{Dillinger:1993,Delabaere19971:exact,Delabaere1999}
\begin{equation}
  \mf{S}_{\arg{\CA}} \CV_\gamma(\uz,\hbar) =
  \CV_\gamma(\uz,\hbar)\left(1+\CV_{\gamma(\CA)}(\uz,\hbar)\right)^{\vev{\gamma,\gamma(\CA)}\SSANS},
\end{equation}
where $\gamma(\CA)$ is the charge associated to $\CA$.  By comparing
with the formula of Kontsevich-Soibelman automorphism, one can
conclude that $\SSANS$ can be identified with BPS invariant
$\Omega(\gamma(\CA))$, in other words
\begin{equation}
  \SSANS = \Omega(\gamma(\CA)).
\end{equation}

\printindex

\bibliographystyle{JHEP}
\bibliography{biblio-stokes}

\end{document}